# Attitudes, willingness, and resources to cover Article Publishing Charges (APC): the influence of age, position, income level country, discipline and open access habits.

Francisco Segado-Boj
Complutense University of Madrid
https://orcid.org/0000-0001-7750-3755

Juan-Jose Prieto-Gutiérrez
COYSODI (Communication and Digital Society) Research Group, Universidad Internacional de la Rioja, Madrid, Spain
https://orcid.org/0000-0002-1730-8621

Juan Martín-Quevedo
Universidad Rey Juan Carlos
https://orcid.org/0000-0003-1005-0469

**Abstract**

The rise of open access (OA) publishing has been followed by the expansion of the Article Publishing Charges (APC) that moves the financial burden of scholarly journal publishing from readers to authors. We introduce the results of an international randomly selected sampled survey (N=3,422) that explores attitudes towards this pay-to-publish or Gold OA model among scholars. We test the predictor role of age, professional position, discipline, and income-level country in this regard. We found that APCs are perceived more as a global threat to Science than a deterrent to personal professional careers. Academics in low and lower-middle income level countries hold the most unfavorable opinions about the APC system. The less experimental disciplines held more negative perceptions of APC compared to STEM and the Life Sciences. Age and access to external funding stood as negative predictors of refusal to pay to publish. Commitment to OA self-archiving predicted the negative global perception of the APC. We conclude that access to external research funds influences the acceptance and the particular perception of the pay to publish model, remarking the economic dimension of the problem and warning about issues in the inequality between center and periphery.

**Key points**
- APCs are perceived more as a global threat to Science than a deterrent to personal professional careers
- Younger academics and those in low and lower-middle income level countries hold the most unfavorable opinions about the APC system.
- The negative perception about APC is higher among Arts & Humanities and Social Science researchers compared to those in STEM and the Life Sciences.
- Age and country income level stood as negative predictors of refusal to pay to publish.

**INTRODUCTION**

Since 2001 Open Access (OA) movement has claimed the need to remove all barriers separating citizens from scientific documents (Prosser, 2004). OA journals are potentially most visible and accordingly, their articles can be more cited (Langham-Putrow, Bakker, & Riegelman, 2021). Also, OA journals have solved partially the financial problems of libraries and institutions that were facing a steep rise in the journal's subscription fees they had to face (Creaser, & White, 2008; Khoo, 2019). According to the Directory of Open Access Journals (DOAJ), in 2021 16,928 have adopted this publishing model.

OA can be reached through two different pathways: Green Road and Gold Road. The former implies the archiving of the manuscript in free-access repositories (Clarke, 2007). The latter requires authors to pay – the so-called Article Processing Charges (APC)- to make their research freely available in the journal's platform (Guédon (2004). In some cases, the OA model has shifted the revenue strategy of publishers, placing their incomes not on the subscribers but the authors.

Only 30% of all OA journals follow an APC model, but 56% of all OA articles are published through this system (Crawford, 2018). Gold OA journals show a wide extent of fees depending, up to some extent, on the nature of the publisher. Titles edited by for-profit publishing houses require higher APCs, as well as those with higher Impact Factor and in medical and science-based disciplines (Siler, Frenken, 2020). Data show that from 2012 to 2018 the average APC cost in the four largest journal publishers rose between 25% and 60% (Khoo, 2020), although the specific determinants to rise the APC appear to vary from publisher to publisher (Asai, 2020). This strained authors' resources and scientific funders, as from 2005 to 2018 the average APC paid by European institutions grew from 858€ to 1,600€, while accounting for inflation it would have risen only to 1,100€ for that period (Aasheim, et al 2019). However, this varies from publisher to publisher and depending on the institution. Cambridge University, for instance, paid in 2018 an average APC of 2,147£ (University of Cambridge, 2018). More recently, Cambridge complained that from 2015 to 2020, more than 30% of the resources of their grants had been paid to Elsevier, averaging 3,302£ per article, and reaching payments of 7,320£ (University of Cambridge, 2021).

Critics have accused this *pay-to-publish model* of translating inequality in the access to knowledge into inequality in the production and diffusion of knowledge, as researchers in low-income countries would lack the funding to cover such APCs and then would be forced to remain consumers of science, not producers of science. For example, an average APC can represent the salary of half a year or more for African authors (Mekonnen, 2021).

Researchers' attitudes towards OA have been identified to be increasingly positive (Joung, Rowley & Sbaffi, 2019; Segado-Boj, Prieto-Gutiérrez, Martín-Quevedo, 2018; Xu et al, 2020; Zhu, 2017). Yet, specific attitude towards APCs has been deemed as 'neutral' to 'somewhat negative' (Tenopir et al, 2017), and seen as the most important downside to OA (Nicholas et al., 2020). Reluctance to Gold OA is deeper in Social Sciences and Humanities than in Life and basic science (Schöpfel et al, 2016; Solomon & Björk, 2016). As traditionally sociohumanistic disciplines are underfunded in comparison to other fields (Solovey, 2020), they may perceive APC as an additional economic burden. Scientists are more willing to pay APCs if they can cover this cost through some research grant rather than with personal funds (Tenopir et al, 2017).

Economic levels also are related to the acceptance of APC. According to Kień́c (2017), researchers from the periphery (from lower income level countries) are more prone to publish in Gold OA journals than those from richer nations. Scientists from lower income countries rely more frequently on their particular budget to pay such APCs than scholars from higher income regions, frequently because of the abovementioned limitations in institutional support.

Different perceptions have also been identified by academic positions. Postdoc researchers have been pointed as more open to the idea of APC than tenured professors or consolidated staff (Tenopir et al, 2017).

**Justification and novelty**

In comparison to the wide array of research on OA in general, the specific topic of APC or Gold OA has not been so widely addressed in the literature recently. Previous surveys have only been addressed to authors subscribed to some publishing house newsletter (Kień́c, 2017), or to specific national communities, be they North America (Tenopir et al, 2017, Halevi & Walsh, 2021), Spain (Ruiz-Pérez & Delgado-López-Cózar, 2017), Brazil (Pavan & Barbosa, 2018) or US/UK (Rowley et al, 2017). We expand such surveys to an international sample of researchers, not restricted by geographical borders or publishing companies. Also, an important drawback is that most of these studies do not take into account variables such as position, age, or discipline (or only tackle one of them), or only do them for researchers from a specific country (Dalton, Tenopir & Björk, 2020).

We adapt factors that have been related to different perceptions of OA (discipline, position, and income-level country) but we also introduce age, as it has been identified as a factor influencing OA attitudes and habits (Eger, Scheufen & Meierrieks 2015; Rodríguez, 2014). We also introduce as a predictor the personal commitment of researchers to OA self-archiving.

Also, we adopt the theory of 'third-person effect', a Media theory positing that people overestimate the effect of media on others compared to their own (Gunther 1991). We compare how do researchers perceive the effect of the APC model on their particular career compared to those in the general scientific ecosystem.

We propose the following objectives:

**1)** Measuring the perception of researchers of the effect of APC on the global ecosystem of science and on their scientific careers.

**2)** Identifying which strategies do authors use to cover APC, including refusal to pay.

**3)** Identifying whether these attitudes and habits (perception of the effect of APC, strategies to cover APCs) can be predicted by age, position, country income level, discipline, access to funds and OA self-archiving practices.

**METHODS**

We extracted the data via a stratified random sampling survey. Our universe of study is corresponding authors of articles published between 2019 and 2020 in Scopus-indexed journals. We retrieved the contact emails via an indirect approach. First, we grouped the existing subject areas in Scopus into four different disciplines. Also, we considered a fifth category with all the journals published in Latin America and Africa to maximise the number of responses from the so-called 'Global South'. Then we randomly selected several journals in each category to reach 95% confidence interval and +/-5% margin error (Table 1 for details). After extracting these journals, we downloaded the information for all the papers they published in the period, including the e-mail information as available in the Scopus database. Collected e-mails were manually screened to eliminate duplicates (Table 1 for sampling procedure details). A final sample of 82,603 identified unique scholars were invited to the survey. We collected 3,422 valid responses (response rate=4.14%) from April 25th to July 10th, 2021.

**TABLE 1.** Sampling details

| Discipline | Aggregated Scopus categories | Sources | Sources (unique) | Sampled journals | Retrieved emails | Emails (unique) |
|---|---|---|---|---|---|---|
| Arts & Humanities (A&H) | 1. Arts and Humanities | 4182 | 3501 | 353 | 6156 | 5955 |
| Life & Health Sciences (L&HS) | 2. Medicine<br>3. Biochemistry, Genetics and Molecular Biology<br>4. Dentistry<br>5. Health Professions<br>6. Immunology and Microbiology<br>7. Neuroscience<br>8. Nursing<br>9. Pharmacology, Toxicology and Pharmaceutics<br>10. Veterinary | 5927 | 4908 | 357 | 21673 | 18395 |
| STEM | 11. Computer Science<br>12. Mathematics e<br>13. Engineering<br>14. Chemical Engineering<br>15. Multidisciplinary<br>16. Chemistry<br>17. Decision Sciences<br>18. Earth and Planetary Sciences<br>19. Energy<br>20. Environmental Science<br>21. Materials Science<br>22. Agricultural and Biological Sciences<br>23. Physics and Astronomy | 14766 | 10112 | 371 | 41640 | 37244 |
| SS (SS) | 24. SS<br>25. Economics, Econometrics and Finance<br>26. Business, Management and Accounting<br>27. Psychology | 11602 | 9685 | 371 | 19422 | 18800 |
| **Africa & Latin America** | | 1199 | 1199 | 292 | 6062 | 4996 |

The form was delivered in English. A Spanish version was offered also to correspondents extracted from Latin American journals. The study design, the form, and the informed consent participants had to agree before accessing the survey were approved by the Institutional Review Board of Universidad Internacional de la Rioja IRB (No PI:004/2021).

**Measurements**

Our survey collected answers from an array of issues related to APCs as well as to the paywall publishing model. The full form is available at: https://doi.org/10.6084/m9.figshare.19121444.v1. We manually aggregated the countries the researchers were working at according to the country income level information in the latest World Bank Report (2021). In the case of Venezuela, this information was last updated in 2019. Given the marginal number of responses from low-income countries, we aggregated low and lower-middle income categories.

We surveyed authors' perception of the APC system and the pay-to-publish model, at both global and particular levels. So, they had to answer through a Likert-scale (1=Not agree at all, 2=Partially disagree, 3=Not agree nor disagree, 4=Partially agree, 5=Totally agree) up to which point they agreed to the sentence 'Globally speaking, the "pay-to-publish" model (requiring Article Processing Charges to authors to publish the accepted papers) damages or slows scientific advancement'.

They also expressed (by the same Likert scale) their agreement to the sentence: 'Personally, speaking from my point of view and personal experience, "pay-to-publish" model (requiring Article Processing Charges to authors to publish the accepted papers) has slowed or damaged my scientific career'. The question warned: 'please, keep in mind that we are now asking about your particular and personal experience, not your general opinion on this regard'.

Other questions asked the participants how they react when considering publishing a manuscript in an APC journal. They had to specify the frequency (1= never, 2=almost never, 3=occasionally, 4=Frequently, 5=always) in which they a) they pay such APC through some research grant, b) their institution covers such APC; c) they pay such APC on their particular or personal budget, d) refuse to publish on APC journals.

We also considered a question regarding practices of open access self-archiving ('How often do you upload your published manuscripts or other research documents to a repository so that they can be freely downloaded by other researchers?') with the Likert scale: 1=Never, 2=Infrequently, 3=Occasionally, 4=Whenever the publisher rights of the journal I submitted the manuscript allows me to do it, 5=Always, even though the publisher rights do not allow me to do it.

**Data analysis**

Results are exposed in medians. Given that our data was ordinal, we chose non-parametric-tests, namely Ordinal Linear Regression (OLR) Models and Kruskal-Wallis Test.

OLR models were used to measure the influence of the considered independent variables (country income level, age group, professional category and OA practices). OLR was chosen as the dependant variables were measured as ordinal values. Tables 4, 5 & 7 show the results of these tests. Among other statistical values, we indicate the global p-value for the overall model. The closer it gets to zero, the most significant it is considered to fit. The traditional threshold of significance stands at p equal to or below .05. We also provide McFadddens $R^2$ which roughly translates as the percentage of variance explained by each model (with 1.0 meaning a perfect fit of 100%). We also provide values for each predictor in the model. Here, again, the lower the p-value, the best each predictor relates to the dependant variable. "Estimate" indicates how many points the outcome changes according to the predictor. Thus, an estimate of 0.25 would indicate that for each point that is added to the predictor, the outcome increases by 0.25 points. A negative estimate denotes a reverse relationship: when the predictor increases, the outcome decreases. SE (standard error) indicates how distant the observed values are from the regression line.

We tested differences among disciplines through Kruskal-Wallis test, considered a standard test for ordinal data (McKight & Najab, 2010). As with the regression model, the closer p-value nears zero, the more certainty can be sustained that the means in the compared groups are statistically significant. Yet, to identify the groups that significantly differ in their means, a posthoc analysis (Dwass-Steel-Critchlow-Fligner) is needed. Two groups are significantly different when the W value is far from zero and the p-value gets closer to zero. To help interpret the differences, we provide the mean differences (md) between each compared group.

We performed all analyses using the R programming language. We used the MASS package (Ripley et al, 2018) for OLR tests.

The following section starts by describing the sociodemographic features of the sample. After, we introduce the disaggregated results for the perception of the effects of APCs on the global and particular levels. Next, we present hypothesis testing (OLR and Kruskal-Wallis) results. Finally, we compare the difference among attitudes by income level country in the four considered disciplines

# RESULTS

The 3,422 participants were mostly older than 36 years and occupied tenured, stable positions. They were affiliated mostly to European and North American institutions and, accordingly, to high income countries. Almost half of them label themselves as STEM researchers (Table 2).

TABLE 2. Sociodemographic features of the participants

| | | n | % |
|---|---|---|---|
| Age | 25 or younger | 35 | 1.0 % |
| | Between 26 and 35 | 704 | 20.6 % |
| | Between 36 and 50 | 1496 | 43.7 % |
| | 51 or older | 1187 | 34.7 % |
| Position | Predoctoral fellow or Ph.D. Student | 455 | 13.3 % |
| | Untenured | 626 | 18.3 % |
| | Tenure-Track | 408 | 11.9 % |
| | Tenured | 1933 | 56.5 % |
| Region (World Bank Classification) | East Asia and Pacific | 335 | 9.8% |
| | Europe and Central Asia | 1428 | 41.7% |
| | Latin America and the Caribbean | 544 | 15.9% |
| | Middle East and North Africa | 140 | 4.1% |
| | North America | 635 | 18.6% |
| | South Asia | 203 | 5.9% |
| | Sub-Saharan Africa | 137 | 4.0% |
| Income level per Capita | High-Income | 2149 | 62.8% |
| | Low Income | 30 | 0.9% |
| | Lower-Middle | 458 | 13.4% |
| | Upper-middle | 785 | 22.9% |
| Disciplines | A&H | 274 | 8.0 % |
| | L&HS | 759 | 22.2 % |
| | SS | 738 | 21.6 % |
| | STEM | 1647 | 48.2 % |

## Perceptions of APCs

The perception of APCs as a general threat to science was higher than it was as detrimental to the particular careers of the participants, as their perception that such a system damages their career is much driven towards indifference (Table 3).

TABLE 3. Global perception of the effect of APCs

| | Global perception | Particular perception |
|---|---|---|
| Global | 4 | 3 |
| 25 or younger | 4 | 3 |
| Between 26 and 35 | 5 | 3 |
| Between 36 and 50 | 4 | 3 |
| 51 or older | 4 | 3 |
| Low and lower middle | 4 | 4 |
| Upper-middle | 5 | 4 |
| High-Income | 4 | 3 |
| Predoctoral fellow or PHD Student | 4 | 3 |
| Untenured | 4 | 3 |
| Tenure-Track | 4 | 3 |
| Tenured | 4 | 3 |
| A&H | 5 | 3 |
| L&HS | 4 | 3 |
| SS | 4 | 3 |
| STEM | 4 | 3 |
| Never | 4 | 3 |
| Infrequently | 4 | 3 |
| Occasionally | 4 | 3 |
| When I am allowed | 4 | 3 |
| Always | 5 | 3 |

Note: The full data with means and standard deviation is available at:
https://doi.org/10.6084/m9.figshare.19102238

By disciplines, the most negative perceptions regarding the global effect of APC were found in the A&H. The Kruskal-Wallis test ($\chi^2$(3, N=3422)= 33.5, p<.001) and the Dwass-Steel-Critchlow-Fligner pairwise comparisons state the existence of such differences in the A&H those in STEM (md=0.31, W=-6.79, p<-001) and the L&HS (md=0.21, W=-4.35, p=.011). Beyond the differences in the median, this statistical test also identified that SS significantly held more adverse opinions than STEM (md=0.21, W=-5.74, p<.001)

The youngest academics tend to express the most unfavorable opinions about the APC system. Likewise, scholars from lower-income level countries express more adverse attitudes (Table 4).

TABLE 4  OLR model for global perception of the effect of APCs

| | | | | Overall Model Test | | |
|---|---|---|---|---|---|---|
| Model | Deviance | AIC | $R^2_{McF}$ | $\chi^2$ | df | p |
| 1 | 8612 | 8632 | 0.0205 | 180 | 6 | <.001 |

| | | Model Coefficients | | |
|---|---|---|---|---|
| Predictor | Estimate | SE | Z | p |

| Predictor | Estimate | SE | Z | p |
|---|---|---|---|---|
| Income | -0.0180 | 0.0459 | -0.392 | 0.695 |
| Position | -0.0396 | 0.0333 | -1.186 | 0.236 |
| Age | -0.2140 | 0.0498 | -4.297 | <.001 |
| Grant | -0.1758 | 0.0243 | -7.223 | <.001 |
| Institution | -0.1804 | 0.0264 | -6.840 | <.001 |
| OA Commitment | 0.0875 | 0.0281 | 3.117 | 0.002 |

Income and position failed to predict global attitudes toward APC publishing, yet OA self-archiving practices positively predicted opinions in this regard. The more researchers self-archived manuscripts, even breaking copyright or publishing agreements, the more hostile their view on the APC model was. Reversely, the more access scholars had to external funding, the more positive their opinions were.

OLR identified that lower income countries and the less consolidated positions were linked to more negative perceptions. Access to research funds also decreased the perceived damage of APCs to researchers' careers. OA self-archiving practices also caused a worse perception of the effect APC had on particular researchers (Table 5).

**TABLE 5.** OLR model for particular perception of the effect of APCs

| | | | | Overall Model Test | | |
|---|---|---|---|---|---|---|
| Model | Deviance | AIC | $R^2_{McF}$ | $\chi^2$ | df | p |
| 1 | 10395 | 10415 | 0.0343 | 369 | 6 | <.001 |

| | | | | Model Coefficients | |
|---|---|---|---|---|---|
| Predictor | Estimate | SE | Z | p | |
| Income | -0.6906 | 0.0444 | -15.56 | <.001 | |
| Position | -0.0994 | 0.0316 | -3.15 | 0.002 | |
| Age | -0.0539 | 0.0472 | -1.14 | 0.253 | |
| Grant | -0.0580 | 0.0234 | -2.48 | 0.013 | |
| Institution | -0.1343 | 0.0255 | -5.27 | <.001 | |
| OA Commitment | 0.0597 | 0.0271 | 2.20 | 0.028 | |

As for the particular perception of APC in the scholar's career, even though the median did not varied by discipline, the Kruskal-Wallis test found a significant difference ($\chi^2$(3,

N=3422)= 20.4, p<.001) in this regard. Dwass-Steel-Critchlow-Fligner pairwise comparisons show that such significant differences were only placed between SS and L&HS (md= 0.09, W=-6.135, p<.001), and STEM (md=0.22, W=-5.1, p<.001).

**Willingness and resources to pay APCs**

Most participants refused to publish in Gold OA journals (Table 6). Still, if they had to pay APCs, they were mostly covered through research grants and less often by the institution itself or the particular budget of the researchers. The oldest researchers paid APC more frequently through some research grant. Scholars from higher income countries also often paid these costs by institutional funds or by research grants and rarely by particular or personal budget. The highest refusal to publish in APC journals was found in lower income countries as well as in A&H and SS.

**TABLE 6.** Medians of strategies to cover APCs

|  |  | By some research grant | By institutional funds | By particular or personal budget | Refuse to publish in APC journals |
|---|---|---|---|---|---|
| Global |  | 2 | 2 | 1 | 3 |
| Age | 25 or younger | 2 | 2 | 1 | 3 |
|  | Between 26 and 35 | 2 | 2 | 1 | 3 |
|  | Between 36 and 50 | 2 | 2 | 1 | 3 |
|  | 51 or older | 3 | 1 | 1 | 3 |
| Income | Low and lower middle | 1 | 1 | 2 | 4 |
|  | Upper-middle | 2 | 1 | 2 | 4 |
|  | High-Income | 3 | 2 | 1 | 3 |
| Position | Predoctoral fellow or PHD Student | 2 | 2 | 1 | 3 |
|  | Untenured | 2 | 2 | 1 | 3 |
|  | Tenure-Track | 2 | 1 | 1 | 3 |
|  | Tenured | 2 | 1 | 1 | 3 |
| Discipline | A&H | 1 | 1 | 1 | 4 |
|  | L&HS | 3 | 2 | 1 | 3 |
|  | SS | 1 | 1.5 | 1 | 4 |
|  | STEM | 3 | 2 | 1 | 3 |
| Commitment to OA | 1 | 2 | 1 | 1 | 3 |
|  | 2 | 2 | 2 | 1 | 3 |
| 3 | 2 | 1 | 1 | 3 | |
| 4 | 2.5 | 2 | 1 | 3 | |
| 5 | 2 | 2 | 1 | 3 | |

Note: The full data with means and standard deviation is available at:
https://doi.org/10.6084/m9.figshare.19102265

The statistical tests confirmed the significance of such differences, as happened with the. Kruskal-Wallis ($\chi^2$(3, N=3422)= 98.1, p<.001). Dwass-Steel-Critchlow-Fligner comparisons rank disciplines almost perfectly in their refusal to APCs from the less experimental to the most experimental: A&H show significantly a more hostile attitude than L&HS (md=0.88, W=-11.92, p<.001), STEM (md=0.54, W=-8.48, p<.001) and social scientists (md=0.32, W=-4.90, p=.003); social scientists also express more rejection to pay to publish than their colleagues from L&HS (md=0.56, W=-10.46, p<.001) and STEM (md=0.22, W=-4.91, p=.003), and STEM are more hostile than Life scientists to publishing in APC journals (md=0.34, W=-7.64, p<.001).

Age and access to external funding stood as negative predictors of refusal to pay to publish. Younger researchers and those lacking institutional support are less willing (or able to) publish in APC journals (Table 7).

TABLE 7. OLR model for refusing to publish in APC journals

| | | | | Overall Model Test | | |
|---|---|---|---|---|---|---|
| | Deviance | AIC | $R^2_{McF}$ | $\chi^2$ | df | p |
| | 9908 | 9928 | 0.0801 | 862 | 6 | <.001 |

| | | | | Model Coefficients | |
|---|---|---|---|---|---|
| Predictor | Estimate | SE | Z | p | |
| Income | -0.0784 | 0.0444 | -1.764 | 0.078 | |
| Position | 0.0271 | 0.0324 | 0.837 | 0.403 | |
| Age | -0.1447 | 0.0482 | -2.999 | 0.003 | |
| Grant | -0.5023 | 0.0250 | -20.073 | <.001 | |
| Institution | -0.4185 | 0.0265 | -15.816 | <.001 | |
| OA Commitment | 0.0378 | 0.0276 | 1.369 | 0.171 | |

In all four disciplines, the perception that the APC model damages the personal career is more common among scholars from lower income level countries. On the contrary, the idea that APC are a global threat to science is more evenly spread among all disciplines and countries of all income level (Table 8).

TABLE 8 Disaggregated Perceptions (median) by discipline and Income-level country

|  |  | Global | Personal | Grant | Institution | Budget | Refuse |
|---|---|---|---|---|---|---|---|
| A&H | High | 5 | 3 | 1 | 1 | 1 | 4 |
|  | Upper-middle | 5 | 3 | 1 | 1 | 1 | 5 |
|  | Low and lower-middle | 5 | 4 | 2 | 1 | 1 | 3 |
| L&H | High | 4 | 3 | 4 | 2 | 1 | 2 |
|  | Upper-middle | 5 | 4 | 2 | 2 | 2 | 3 |
|  | Low and lower-middle | 4 | 4 | 1 | 1 | 2 | 4 |
| SS | High | 4 | 2 | 1 | 2 | 1 | 3 |
|  | Upper-middle | 5 | 3 | 1 | 1 | 1 | 4 |
|  | Low and lower-middle | 4 | 3 | 2 | 1 | 2 | 3 |
| STEM | High | 4 | 3 | 3 | 2 | 1 | 3 |
|  | Upper-middle | 4 | 4 | 2 | 1 | 1 | 4 |
|  | Low and lower-middle | 5 | 4 | 1 | 1 | 1 | 4 |

Refusal to publish in APC models seems linked to income-level in the most experimental disciplines (STEM & L&HS) where the median for refusing APC-based journals increases as the income level decreases. Something similar happens with access to funds. L&H and STEM scientists pay more often APCs via research grants or institutional funds when their institutions are based in higher income level countries. In L&HS it is more common in the lower countries to use their personal budget. Yet, these patterns regarding how APCs are paid are absent in A&H and SS.

## DISCUSSION AND CONCLUSIONS

APCs rise more concern as a global threat to scientific development than as a hindrance that personally affects the career of the participants.
The perception that APCs are detrimental to science in a global level might be related to other facets of Gold OA, such as the quality of the peer-review process and the quality of the accepted and published results (Ruiz-Pérez & Delgado-López-Cózar, 2017), to the rise of predatory publishing (Al-Khatib & Teixeira Da Silva, 2017) or to an understanding that it aggravates previous inequalities among scholars from different economic backgrounds.

While Kień (2017) and Pavan & Barbosa (2018) stated that researchers from lower income countries opted more frequently for Gold OA journals, our results provided opposing evidence. Scientists from the economic periphery reflected worse opinions to APC (both globally and individually) than scientists from middle and high income countries. Additionally, they also rejected this editorial model more often.

This difference may be explained because while Kień used a sample from the distribution list of a commercial publishing house, we invited our contestants through a random approach. Even it may be plausible that the universe of Kień was more familiar with Gold OA journals and thus, less hostile to the APC model. As for Pavan & Barbosa, they analysed actually published documents. Their data information might have missed the attitudes of authors who in fact could not or chose not to publish in APC-funded journals.

This rejection of APCs from lower income countries might be related to the scarcity of economic capital or external funds that damages the trajectories of scholars from non-Western countries (Goyanes and Demeter, 2021). Krauskopf (2021) studies the case of Chile, pointing out that the structural trend of lacking grants and funds stops researchers from being able to publish in high-impact, APC charging journals. Thus, this may explain why scientists from such countries might see APCs and the Gold OA model as another, heavier, burden, especially as some studies point out that up to 60% of the researchers in low-income countries have to pay the APC by themselves (Nobes & Harris, 2021).

Lack of economic capital might as well explain also why lower professional positions predict more negative views of the APCs on a particular level. While youth might, at least theoretically, imply greater openness to alternative models of scholarly publishing. Our data imply that the generational factor is subdued to structural constraints in the professional track. Researchers in the first stages of their careers typically have less financial aid and thus might feel damaged by the APC model and are less willing to follow the Gold road to OA. This contradicts partially previous findings (Tenopir, 2017) that those in the early stages of their careers held a more positive view of Golden OA. Once again, we think that differences in the sample would explain this contradiction. Tenopir and colleagues restricted their survey to staff in four North-American intensive research universities, while our participants reflect a wider, more diverse, array of institutions located around the world.

We found a generational gap. Younger academics are more critical of the global effects of the pay-to-publish paradigm. The youngest scholars' initial reluctance is deeper in the 26-35 cohort. We interpret that as ECRs get familiar with the full scientific ecosystem – beyond their particular lab- they are aware of the structural consequences of the APC model, deepening their initial criticism.

As for disciplines, rejection to APCs is higher in A&H and SS than in STEM and L&HS, as previously seen (Schöpfel et al, 2016, Tenopir, 2017). While some authors argue that the familiarity with OA, in general, reflects in openness to Gold OA (Schöpfel et al, 2016), others have reasoned that the availability of grants explains why L&HS or STEM are more willing to pay APC (Tenopir, 2017).

In the light of our findings, we find that rejection for APCs is likely driven by economic reasons, and that these tend to outweigh cultural or personal circumstances that would lead to a more favorable view. F.i., OA self-archiving habits played no effect on the refusal to publish in APC charging journal. The main predictor in the perception and attitudes to the *pay-to-publish* model is access to external funding. Those who hardly can acquire external funds for their research are less keen on a publishing model that places more expenses on their shoulders.

'Relative wealth' of disciplines, understood as the funding granted to research in each field, has been identified as a factor explaining the preference of L&HS and STEM to Gold OA (Björk et al., 2010, Boukacem-Zeghmouri et al, 2018). This 'relative wealth' is also a factor that explains the rejection of Golden OA in the cases of ECR and lower income countries. We face the problem that, while the lack of grants hinders, but not makes unfeasible, research in the SS and the Humanities, the expansion of the APC model might make it effectively impossible to communicate the results of this research. This problem may be even worse in the lower income level countries, deepening current issues of inequality between center and periphery

As for limitations, as we restricted the sample to corresponding authors, it might be biased towards more consolidated researchers to detriment of ECRs. Also, as we retrieved our sample from the Scopus database, our data might be extrapolated only to researchers publishing in indexed journals. We did not include professional experience as a predictor in our models, but we believe that professional situations can be taken as an indirect proxy for professional experience. The form did also not include the option for respondents to state that they were currently unemployed, so affiliation information might not be absolutely accurate.

Our results open the question of why the effects of APCs are deemed worse for science as a whole than for particular researchers. Further research is needed to better seize the different aspects of how and why scholars perceive that APCs might damage the advancement of science. As stated, the economic difficulties derived from Gold OA to publish the research results, are only one of the aspects where APCs might be prejudicial. Future surveys could measure aspects related to the quality of peer review or other questions.

**Conflict of interest**

The authors have no conflict of interest relevant to this article

**Contributions**: FSB conceived the project, FSB, JMQ and JJPG developed the methodology, JJPG and JMQ collected the data, FSB analysed the data, FSB, JJPG & JMQ wrote the article.